\documentclass[aps,prl,twocolumn,floatfix,superscriptaddress,longbibliography]{revtex4-2}
  
\usepackage{braket}
\usepackage{bbold}
\usepackage{hyperref}
\usepackage{subfigure}
\usepackage[usenames,dvipsnames]{color}
\usepackage{subfigure}
\usepackage[mathscr]{euscript}
 
\usepackage[english]{babel}
\usepackage{epsfig,amsopn}
\usepackage{graphicx}
\usepackage{color}
\usepackage{amsmath,amssymb}
 \usepackage{braket}
\usepackage{relsize}
\usepackage{comment}
\usepackage{enumerate}
\newcommand\bea{\begin{eqnarray}}
\newcommand\eea{\end{eqnarray}}
\newcommand\beq{\begin{equation}}
\newcommand\eeq{\end{equation}}
\newcommand{\noi}{\noindent}

\def\nn{\nonumber}
\def\f{\frac}

\def\de{\delta}

\def\ga{\gamma}

\def\ra{\rangle}

\begin{document}
\title{All-electrical scheme for valley polarization in graphene} 
 \author{Sachchidanand Das}
 \affiliation{School of Physics, University of Hyderabad, Prof. C. R. Rao Road, Gachibowli, Hyderabad-500046, India}

  \author{ Abhiram Soori}
  \email{abhirams@uohyd.ac.in}
  \affiliation{School of Physics, University of Hyderabad, Prof. C. R. Rao Road, Gachibowli, Hyderabad-500046, India}
\begin{abstract}
We propose an all-electrical setup to generate valley polarization in graphene. A finite graphene sheet is connected to two normal metal electrodes each with two terminals along its zigzag edges, while the armchair edges remain free. When a bias is applied to one terminal and the others are grounded, valley polarization emerges due to transverse momentum matching between the graphene and the metal electrodes. Significant valley polarization is achieved when the Fermi wavevector in the metal exceeds half the separation between the \( K \) and \( K' \) valleys in graphene. We analyze how conductance and valley polarization depend on geometric and electronic parameters. While increasing the width enhances both conductance and polarization, increasing the length introduces Fabry--P\'erot oscillations and suppresses valley polarization due to enhanced intervalley mixing. We also examine the effects of disorder: on-site disorder in graphene increases conductance near the Dirac point  but reduces valley polarization. Finally, we study the impact of imperfect armchair edges and interface roughness, finding that moderate deviations from ideal conditions still yield substantial valley polarization. Our results demonstrate a viable route to electrically controlling valley degrees of freedom in graphene-based devices.
\end{abstract}
\maketitle
{\it Introduction.--}
Graphene, a two-dimensional material composed of carbon atoms on a honeycomb lattice, has attracted significant attention due to its unusual band structure and associated quantum phenomena such as Klein tunneling~\cite{Castro08}. Its valence and conduction bands meet at the Dirac points with linear dispersion. These Dirac points, located at inequivalent momenta labeled \( K \) and \( K' \), define graphene’s two valleys—an internal quantum degree of freedom that can be selectively manipulated. This forms the basis of valleytronics, which exploits valley polarization (an imbalance in populations at \( K \) and \( K' \)) as an information carrier~\cite{Schaibley16}. The valley degree of freedom offers new possibilities for applications in electronics, quantum computing~\cite{AlonsoCalafell19}, optoelectronics~\cite{WANG19}, and energy-efficient devices.

Several approaches to achieving valley polarization in graphene have been proposed. Strain and symmetry-breaking potentials in graphene quantum dots can generate valley polarization~\cite{Li20}, as can gate-controlled \( n \)--\( p \)--\( n \) transistors~\cite{Garcia-Pomar08} and graphene-superconductor junctions~\cite{Akhmerov07}. In biased bilayer graphene, a band gap allows circularly polarized light to produce valley polarization~\cite{Friedlan13}. Terahertz radiation~\cite{Abergel09,Sharma25} and other light-matter interactions~\cite{Mrudul21,Rana23,Mrudul2021} have also been shown to filter electrons by valley.

Fewer proposals address valley polarization through purely electrical means. Luo et al.~\cite{Luo17} studied a point contact with a superconducting lead; valley filtering was also found in junctions between zigzag nanoribbons of differing widths~\cite{Li21}, and in three-terminal devices with mixed edge terminations~\cite{Zhang21}. All-electrical control in bilayer graphene has also been reported~\cite{Shimazaki15,Chen22}. However, valley polarization at interfaces between graphene and conventional metals remains unexplored. Our work addresses this gap by proposing a conceptually simple and experimentally feasible setup using normal-metal electrodes to generate valley polarization in monolayer graphene.

\begin{figure} [htb]
    \includegraphics[width=0.9\linewidth]{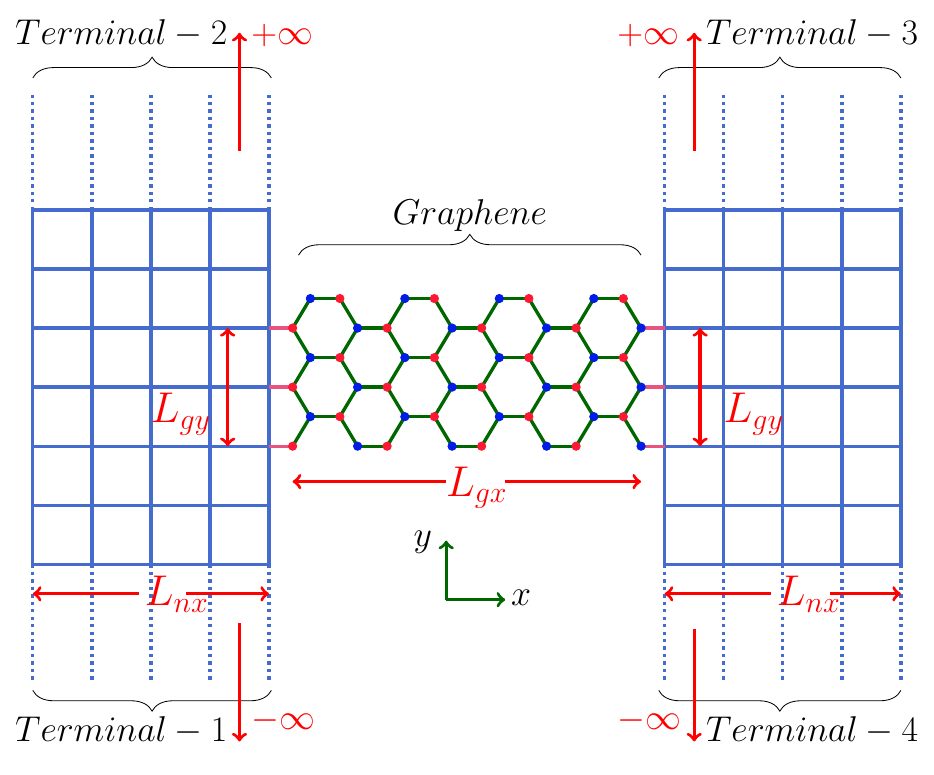}
    \caption{Schematic of the normal metal/graphene/normal metal junction wherein valley polarization can be acheived.}
    \label{fig:schem}
\end{figure}

{\it The System and the Central Idea.--}
Electron transmission across a junction between two-dimensional materials occurs when transverse momentum is conserved~\cite{suri21,soori2021,Sahoo2023}. Based on this, we propose a setup for generating valley polarization in graphene using a purely electrical method. A finite graphene sheet is placed between two square-lattice normal metals, as shown in Fig.~\ref{fig:schem}. The graphene is oriented such that its zigzag edges contact the metal electrodes, while the armchair edges remain free. A bias is applied at terminal 1, with terminals 2, 3, and 4 grounded.

On the biased metal side, the current flow along the $y$-direction leads to an asymmetric occupation of $k_y$ states. If the valley separation in graphene exceeds $2k_{y,F}$, where $k_{y,F}$ is the Fermi wavevector of the metal in the $y$-direction, transmission occurs primarily into one valley, producing valley polarization. Though the system lacks full translational invariance, this momentum filtering remains valid for sufficiently wide graphene regions. We also find that while disorder enhances conductance near the Dirac point, it suppresses valley polarization.

The graphene region has dimensions $L_{gx} \times L_{gy}$, and the metal electrodes extend infinitely along $y$ and have finite width $L_{nx}$ along $x$. Nearest-neighbor hopping in graphene is $\gamma$, and in the metal, it is $t$. The chemical potentials in graphene and metal are $\mu_g$ and $\mu_n$, with $\mu_g$ tunable via gate voltage.

We compute the differential conductance $G_j = dI_j/dV_1$, where $I_j$ is the current in terminal $j$, $V_1$ is the applied voltage at terminal 1, and all other terminals are grounded. The conductance through graphene is $G_g = G_1 - G_2$. These are calculated using Landauer-B\"uttiker scattering theory.

We further resolve the conductance into valley-resolved components, $G_{g,K}$ and $G_{g,K'}$, and define the valley polarization efficiency as
$\eta = {2(G_{g,K} - G_{g,K'})}/{(G_{g,K} + G_{g,K'})}$. This quantifies the extent of valley polarization.

\begin{figure} [htb]
          \includegraphics[width=6.5cm]{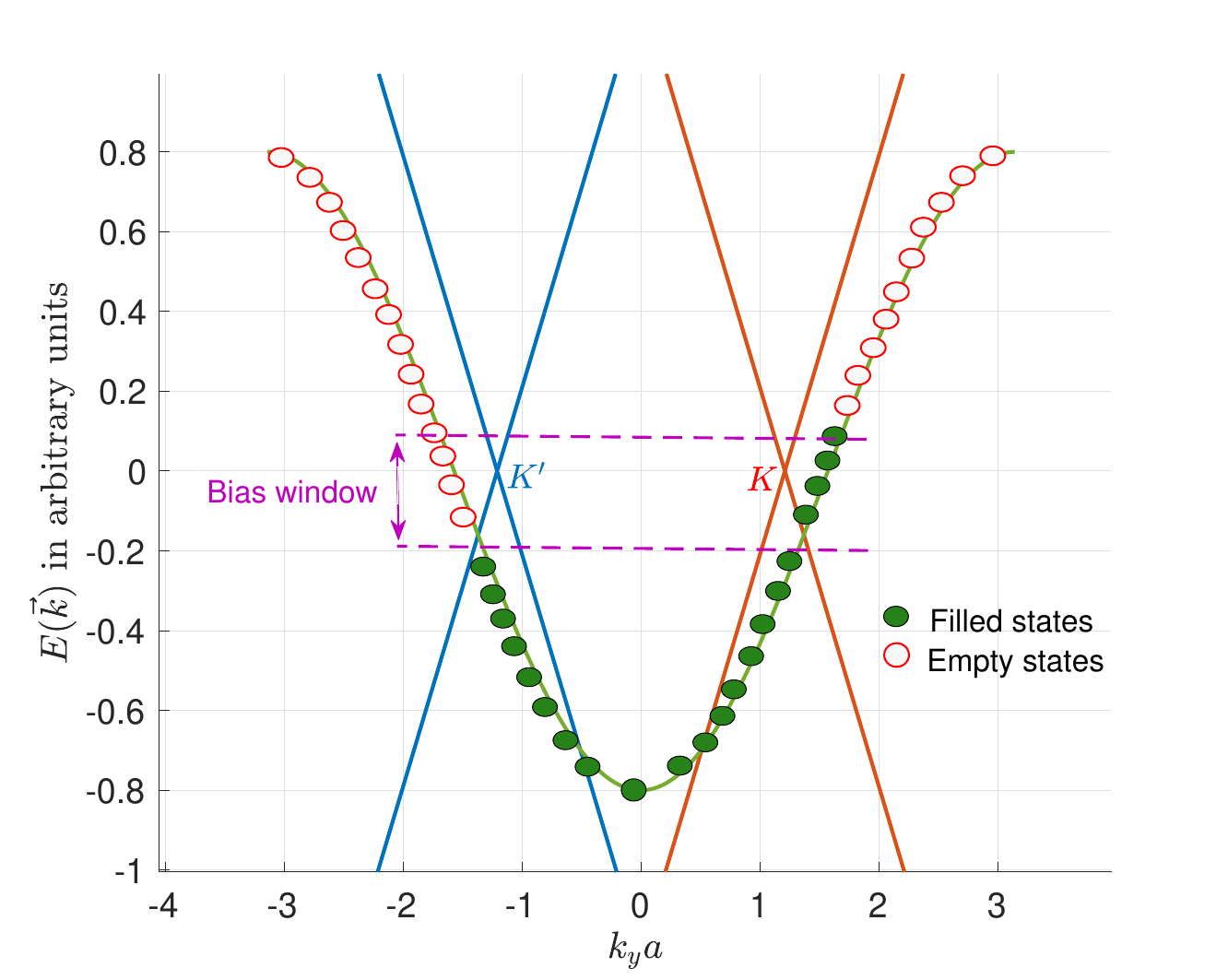}
          \caption{Dispersion of graphene and normal metal fixing $k_x=0$. }
          \label{fig:disp}
    \end{figure}

The mechanism is illustrated in Fig.~\ref{fig:disp}, where $E$ vs $k_y$ for $k_x = 0$ shows that within the bias window, $k_y$ values in graphene and metal align, enabling valley-selective transmission.

\begin{figure}[htb]
 \includegraphics[width=8.0cm]{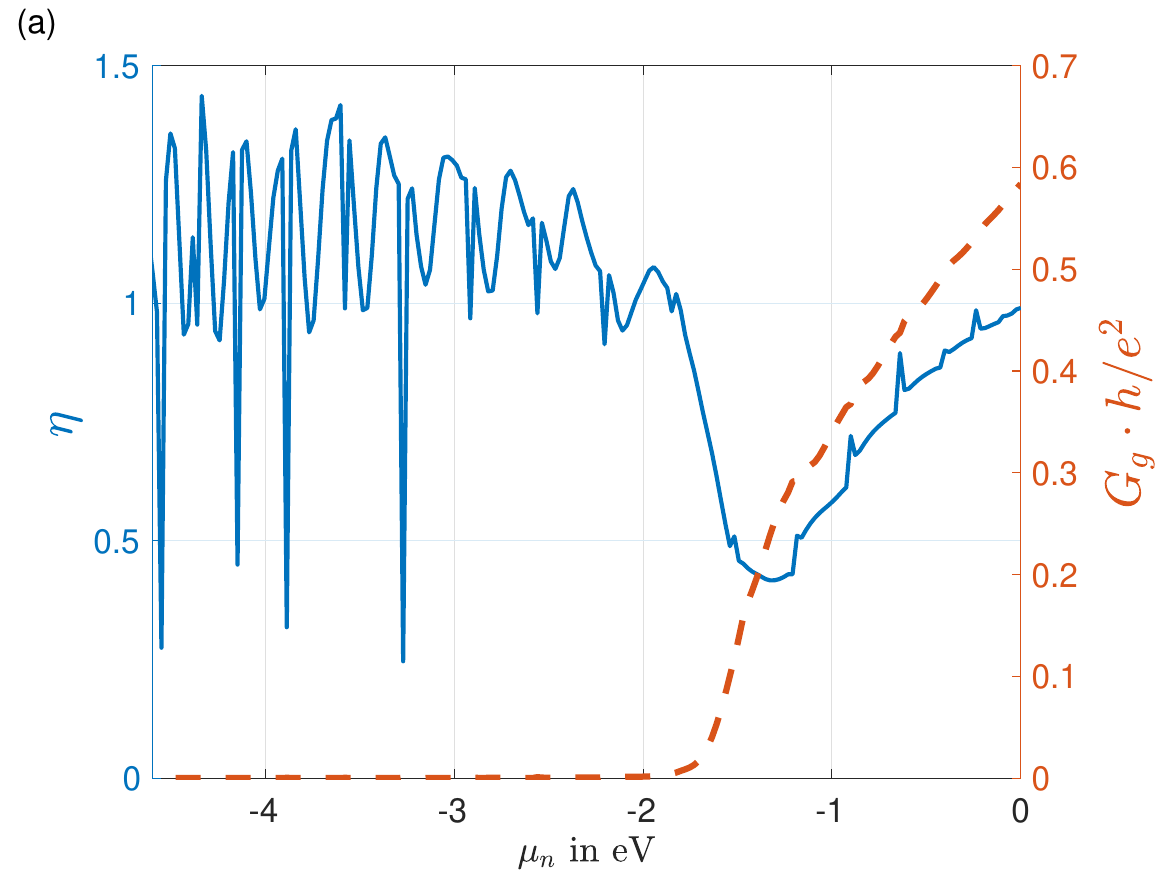}
 \includegraphics[width=4.0cm]{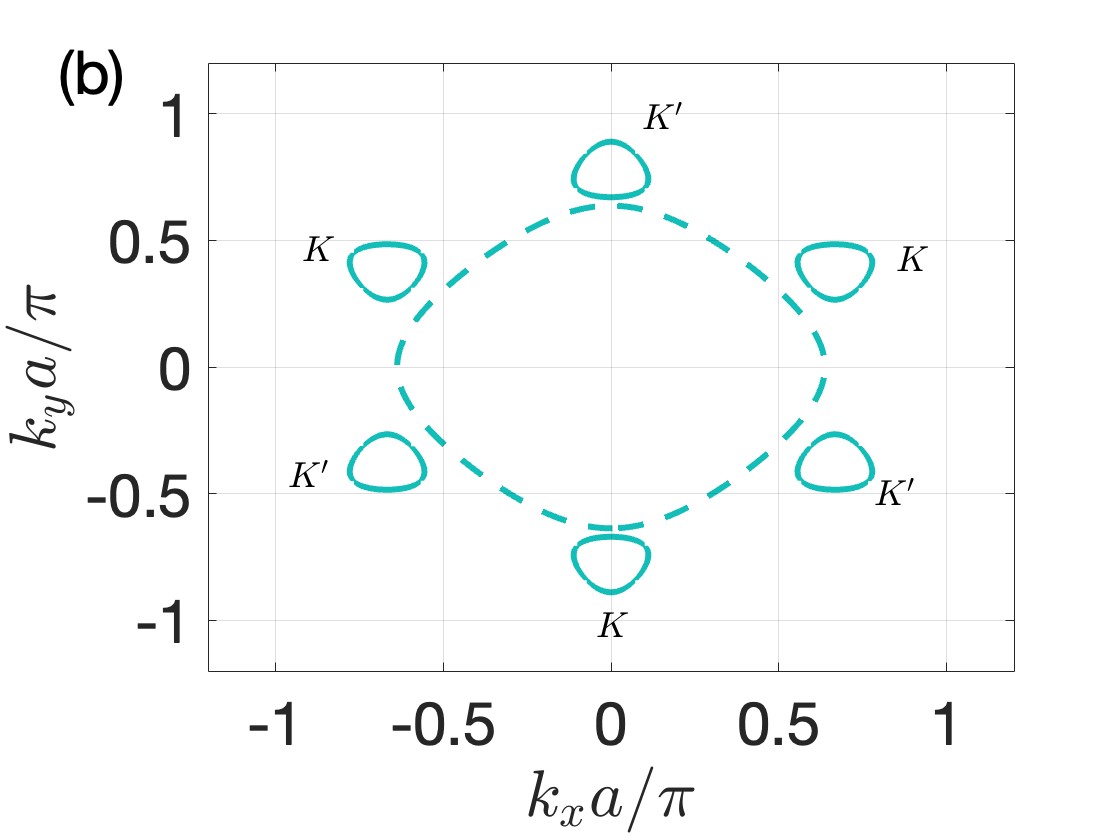}
 \includegraphics[width=4.0cm]{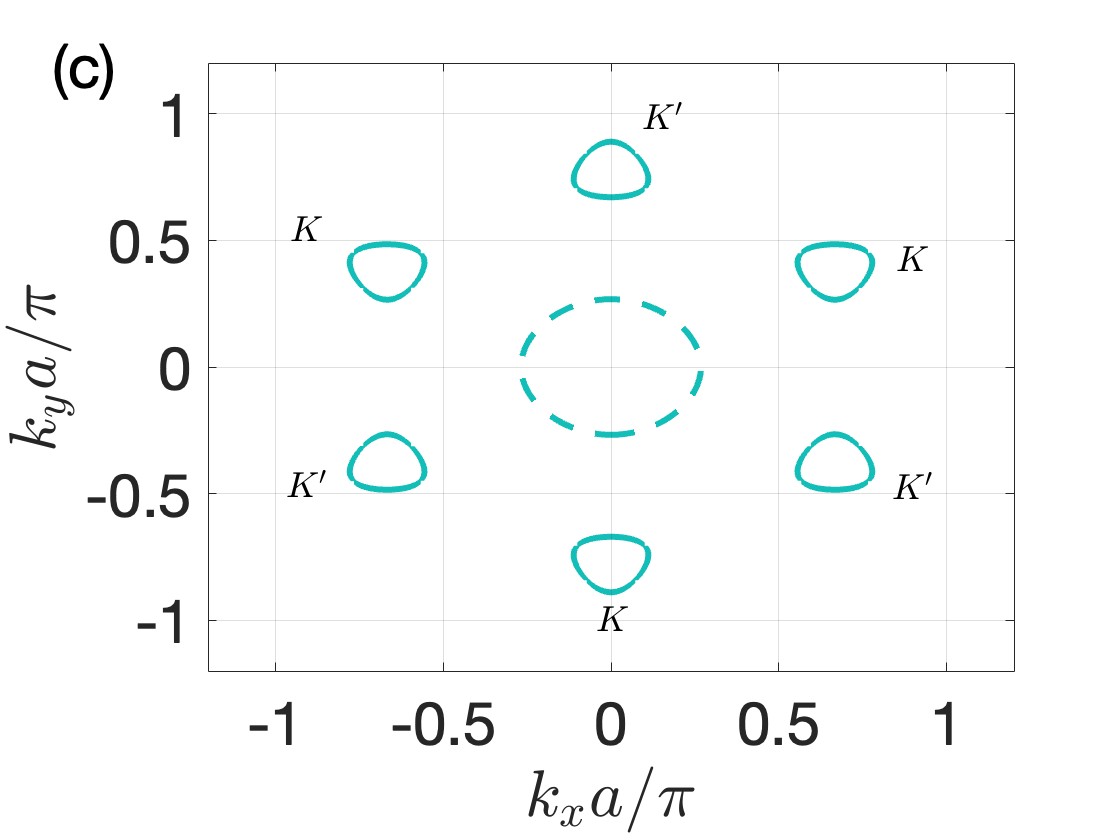}
    \caption{(a) Conductance through graphene (right axis) and valley polarization efficiency (left axis) versus $\mu_n$ at zero bias. Parameters:  $t=1.2eV$, $L_{nx}=20$, $L_{gy}=24$, $L_{gx}=24$, $\mu_g=0.5eV$, $\ga=2.7eV$. The Fermi surfaces of normal metal (dashed line) and graphene (solid line) for (b) $\mu=-1.4eV$ (c) $\mu=-4eV$. }
 \label{fig:mu}
 \end{figure} 
 
  {\it Results.--} 
Varying the normal metal’s chemical potential $\mu_n$ alters its Fermi surface size. Figure~\ref{fig:mu}(a) shows the graphene conductance $G_g$ and valley polarization efficiency $\eta$ versus $\mu_n$. Panels (b) and (c) display Fermi surfaces for $\mu = -1.4~\text{eV}$ and $-4~\text{eV}$, respectively. At $\mu = -1.4~\text{eV}$, the Fermi surfaces of graphene and the metal share a broader $k_y$ range, resulting in higher conductance. At $\mu = -4~\text{eV}$, the $k_y$ mismatch lowers $G_g$, as also reflected in Fig.~\ref{fig:mu}(a), where $G_g$ is nearly zero for $\mu_n \lesssim -1.8~\text{eV}$, and most electrons from terminal-1 reach terminal-2.

Despite the low $G_g$ at $\mu = -4~\text{eV}$, $\eta$ remains high because states in the $K$ valley are closer to the positive $k_y$ states of the metal than those in $K'$. Near $\mu = -1.5~\text{eV}$, $\eta$ drops as $K$ valley states with high $v_y$ undergo repeated reflections in graphene along $y$, mixing with $K'$ valley states. These reflections reduce $\eta$ by increasing $G_g$ without proportionally enhancing $G_{g,K} - G_{g,K'}$.

 \begin{figure}[htb]
 \includegraphics[width=4.0cm,height=3.5cm]{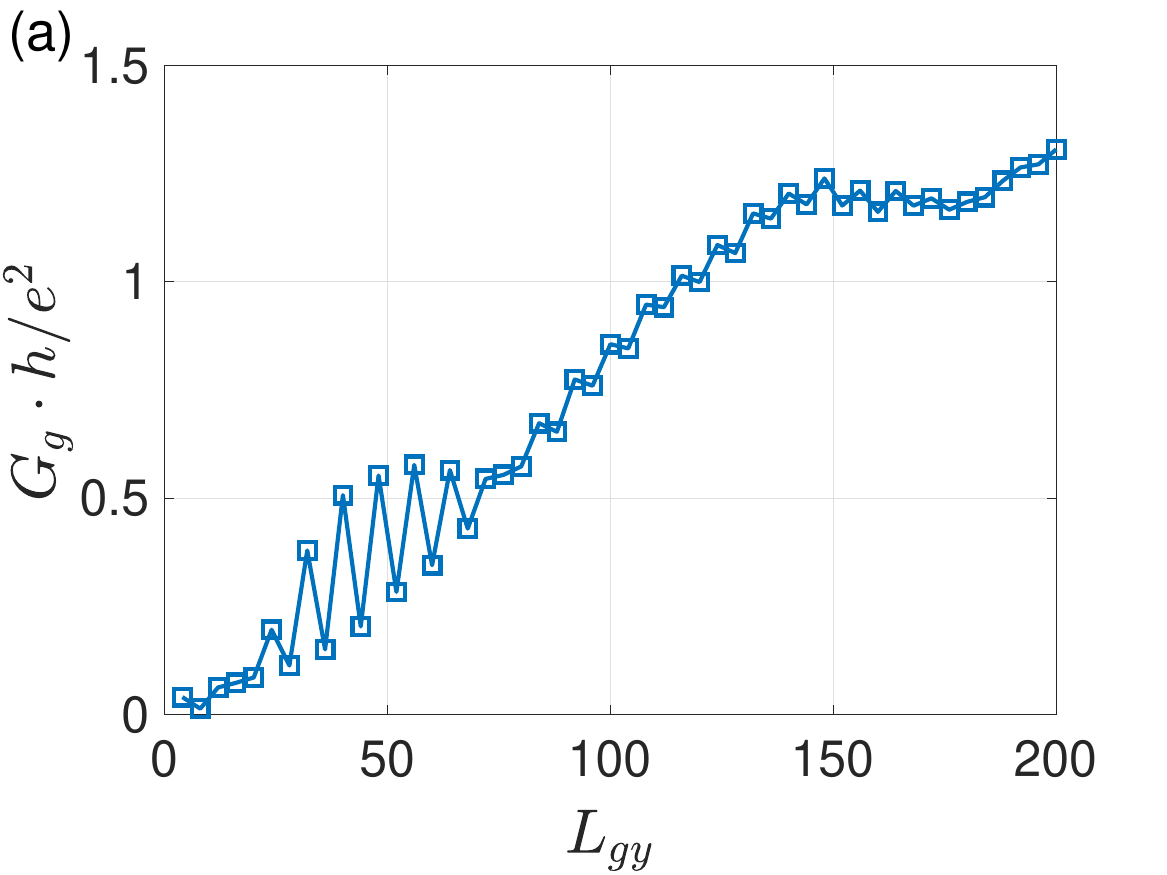}
 \includegraphics[width=4.0cm,height=3.5cm]{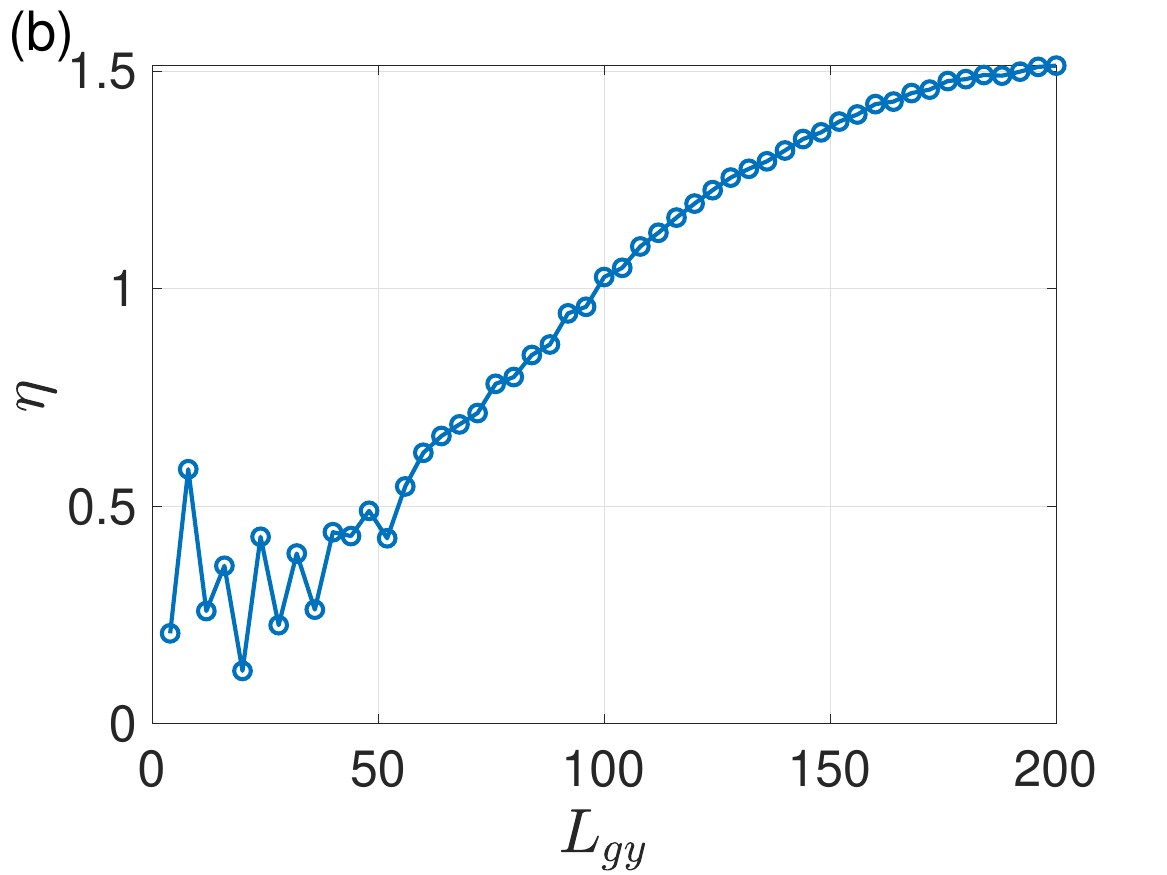}
 \includegraphics[width=4.0cm,height=3.5cm]{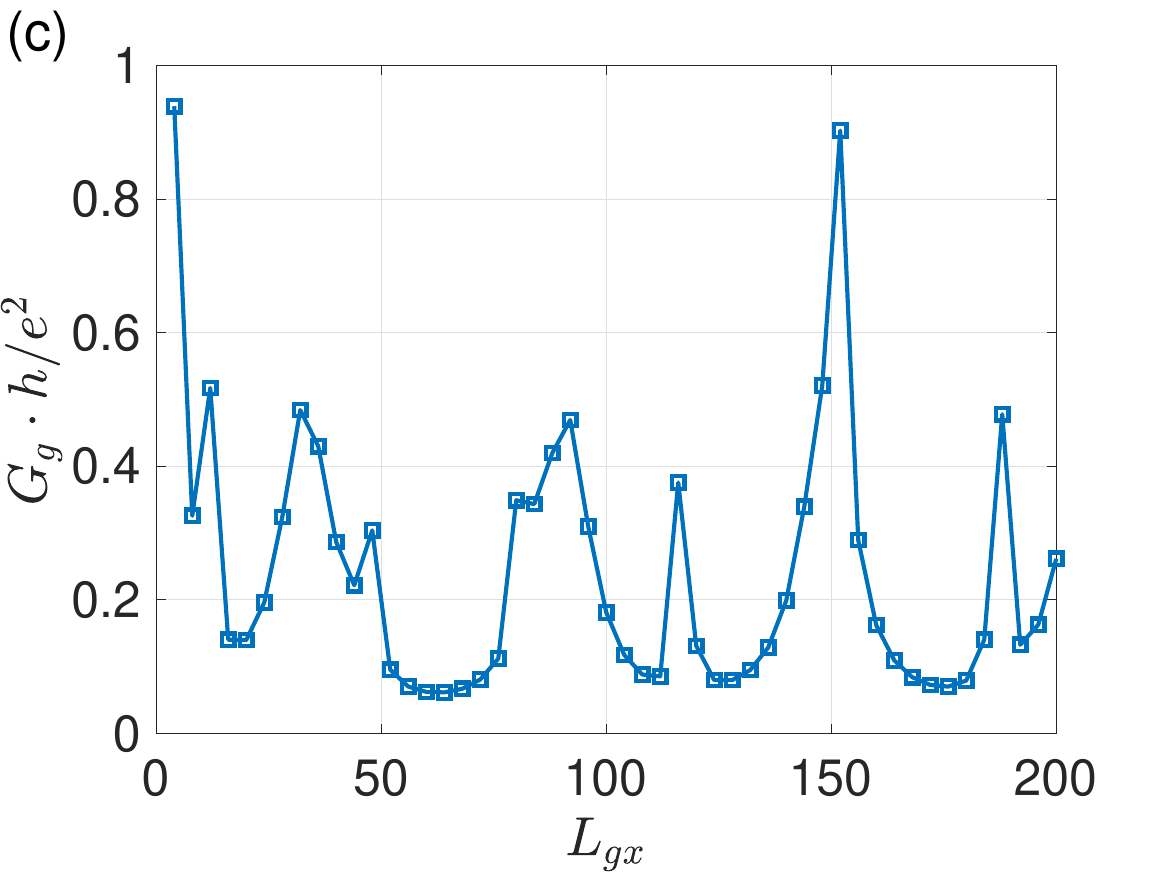}
 \includegraphics[width=4.0cm,height=3.5cm]{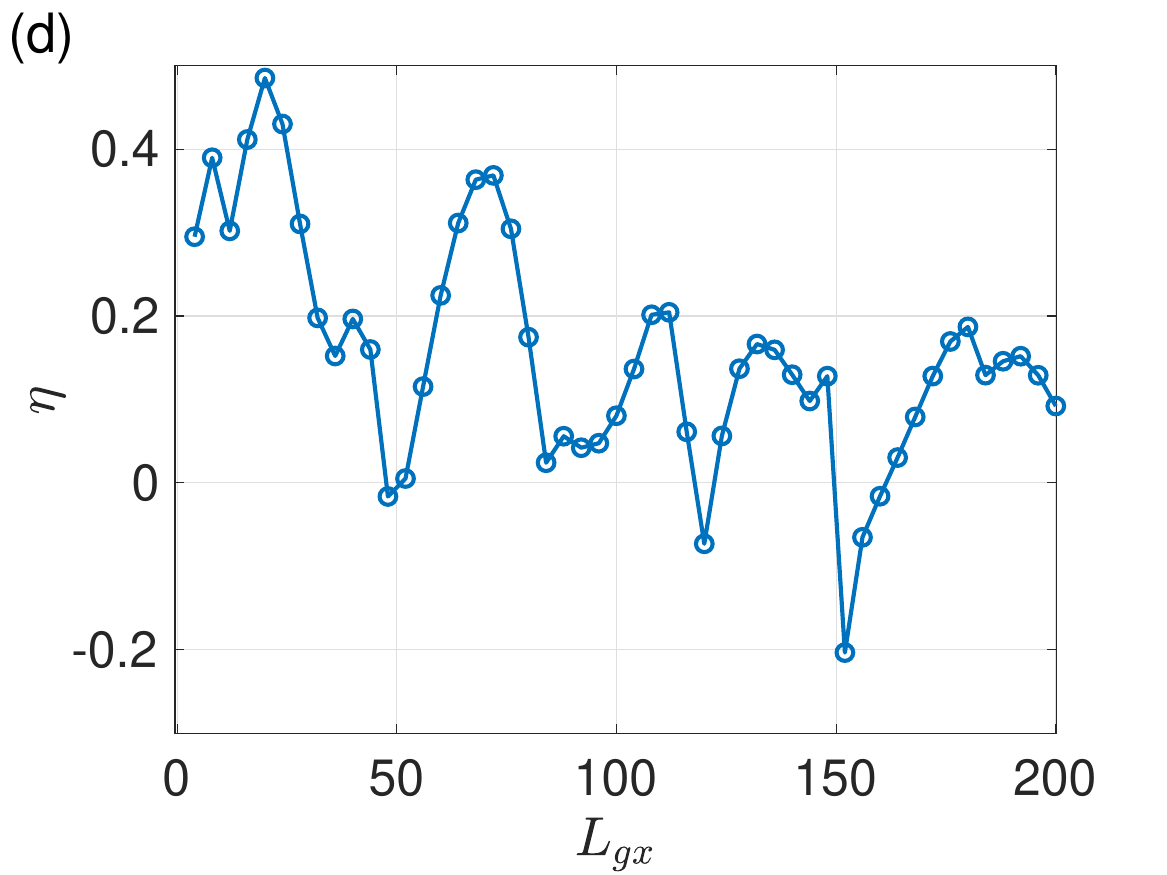}
    \caption{(a) Conductance through graphene $G_g$ versus $L_{gy}$, (b) Valley polarization effeiciency $\eta$ versus $L_{gy}$, (c) $G_g$ versus $L_{gx}$, (d) $\eta$ versus $L_{gx}$. Parameters: (a,b) $L_{gx}=24$, (c,d) $L_{gy}=24$.  $t=1.2~eV$,  $\mu=-1.4~eV$, $\mu_g=0.5~eV$. }
    \label{fig:LgxLgy}
\end{figure}

This mechanism is confirmed by studying $G_g$ and $\eta$ versus $L_{gx}$ and $L_{gy}$. Increasing $L_{gy}$ (fixed $L_{gx}$) boosts both $G_g$ and $\eta$, as shown in Fig.~\ref{fig:LgxLgy}(a,b). In contrast, increasing $L_{gx}$ (fixed $L_{gy}$) initially suppresses $G_g$, which then oscillates [Fig.~\ref{fig:LgxLgy}(c)], while $\eta$ decreases overall [Fig.~\ref{fig:LgxLgy}(d)]. These trends support the role of transverse reflections in valley mixing.

Among conduction channels in graphene, some lie outside the bias window and transmit via evanescent modes, causing $G_g$ to decay with increasing $L_{gx}$. Others transmit via plane-wave modes, yielding oscillatory conductance. The combined result is an initial drop in $G_g$ followed by irregular oscillations—which become regular at smaller $L_{gy}$ (see Supplementary Material for detailed discussion). Meanwhile, $\eta$ decreases with $L_{gx}$ due to enhanced back-and-forth scattering in the $y$-direction.

\begin{figure}
\includegraphics[width=8cm]{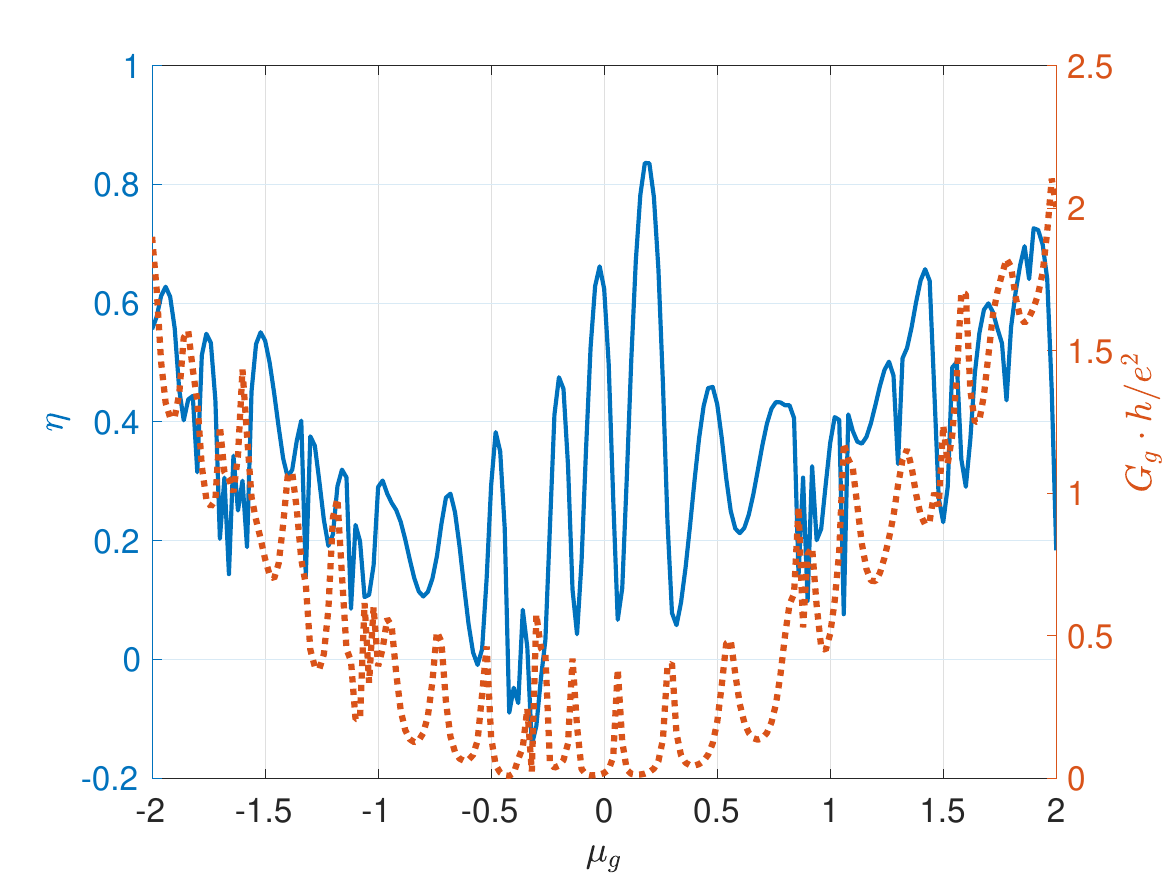}
\caption{Conductance $G_g$ and valley polarization efficiency $\eta$ versus the chemical potential of graphene $\mu_g$. Parameters: $\mu=-1.4~eV$, $L_{gx}=L_{gy}=24$, $L_{nx}=20$. }\label{fig:mug}
\end{figure}

We next examine how $G_g$ and $\eta$ vary with the graphene chemical potential $\mu_g$ [Fig.~\ref{fig:mug}]. As $\mu_g \to 0$ (Dirac point), $G_g$ drops to zero, accompanied by oscillations. This vanishing conductance arises from the low density of states near the Dirac point. Moving away from $\mu_g = 0$, the increasing density of states enhances $G_g$, with oscillations due to Fabry-P\'erot interference of plane-wave modes~\cite{soori22car,suri21,soori2021}. The valley polarization efficiency $\eta$ peaks near $\mu_g = 0$, but since $G_g$ is close to zero there, the polarization is of limited practical relevance.

\begin{figure}
\includegraphics[width=4cm]{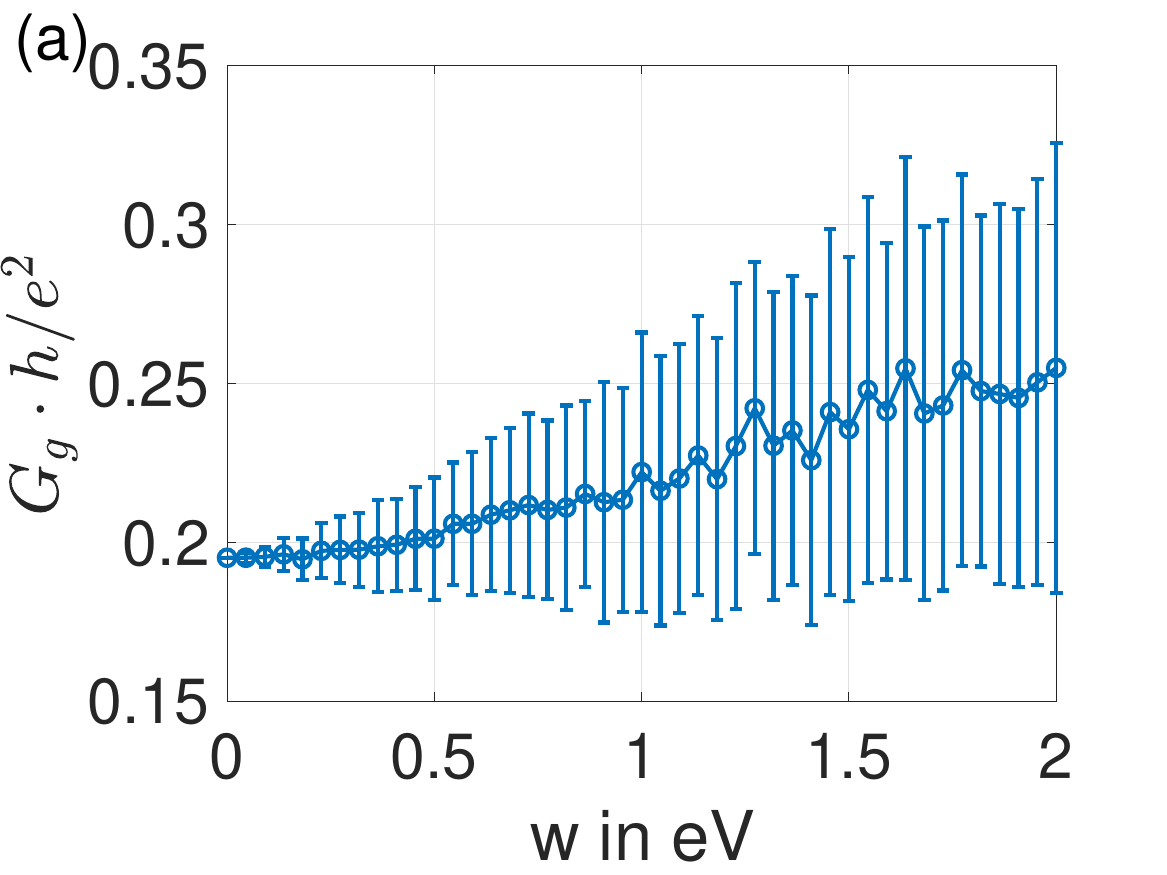}
\includegraphics[width=4cm]{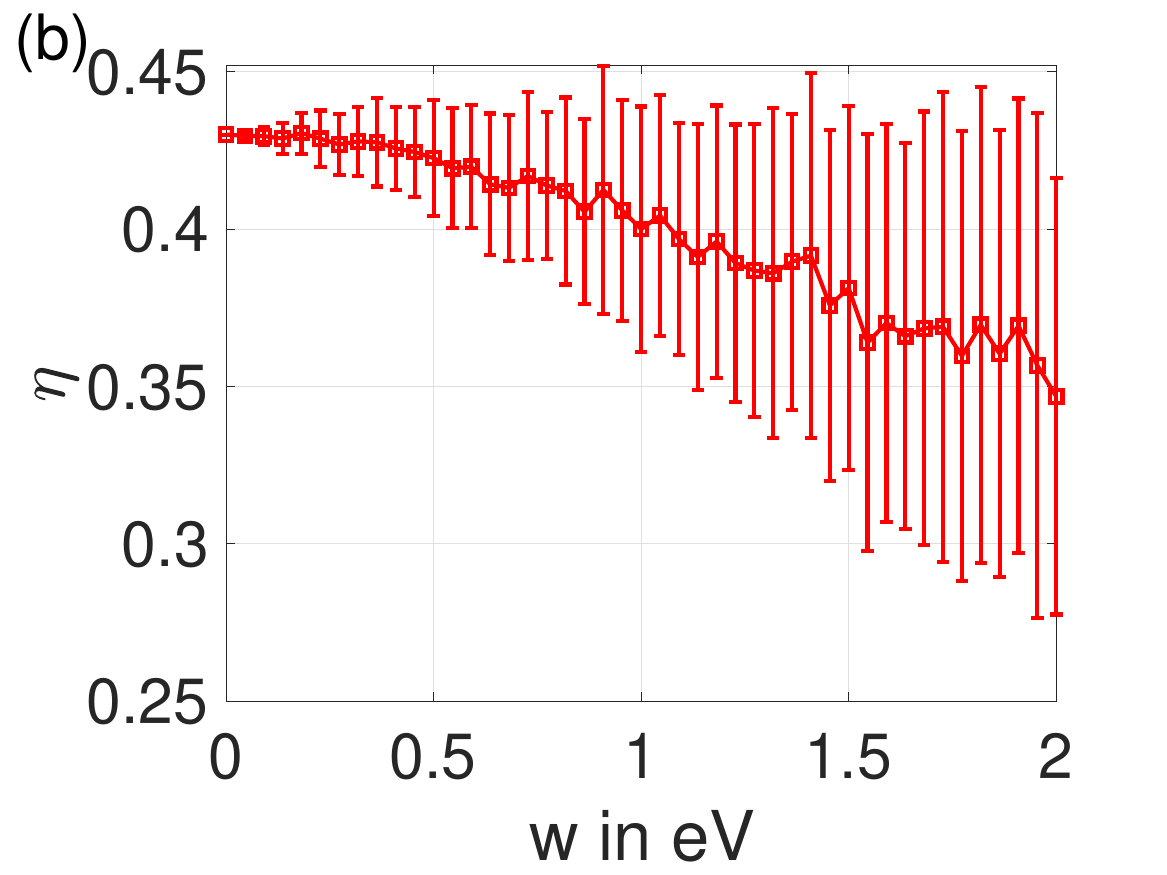}
\includegraphics[width=4cm]{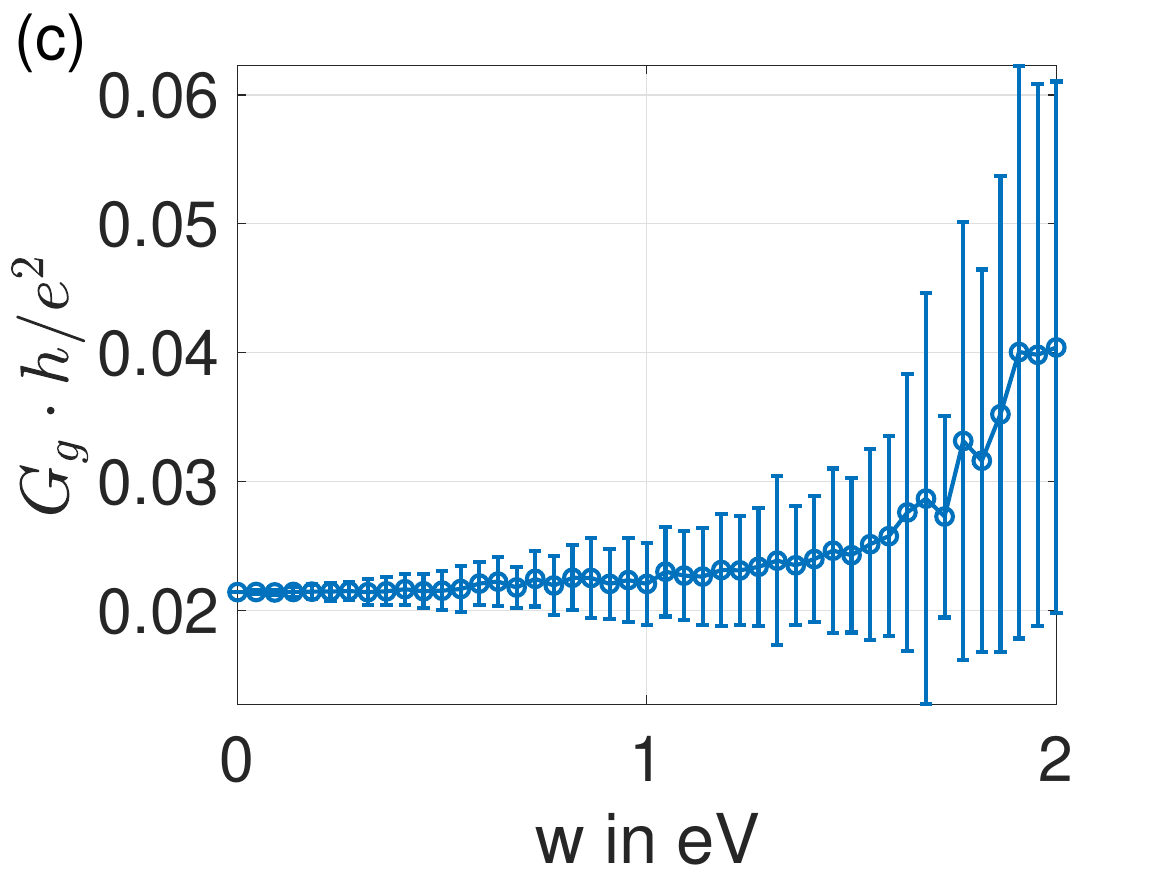}
\includegraphics[width=4cm]{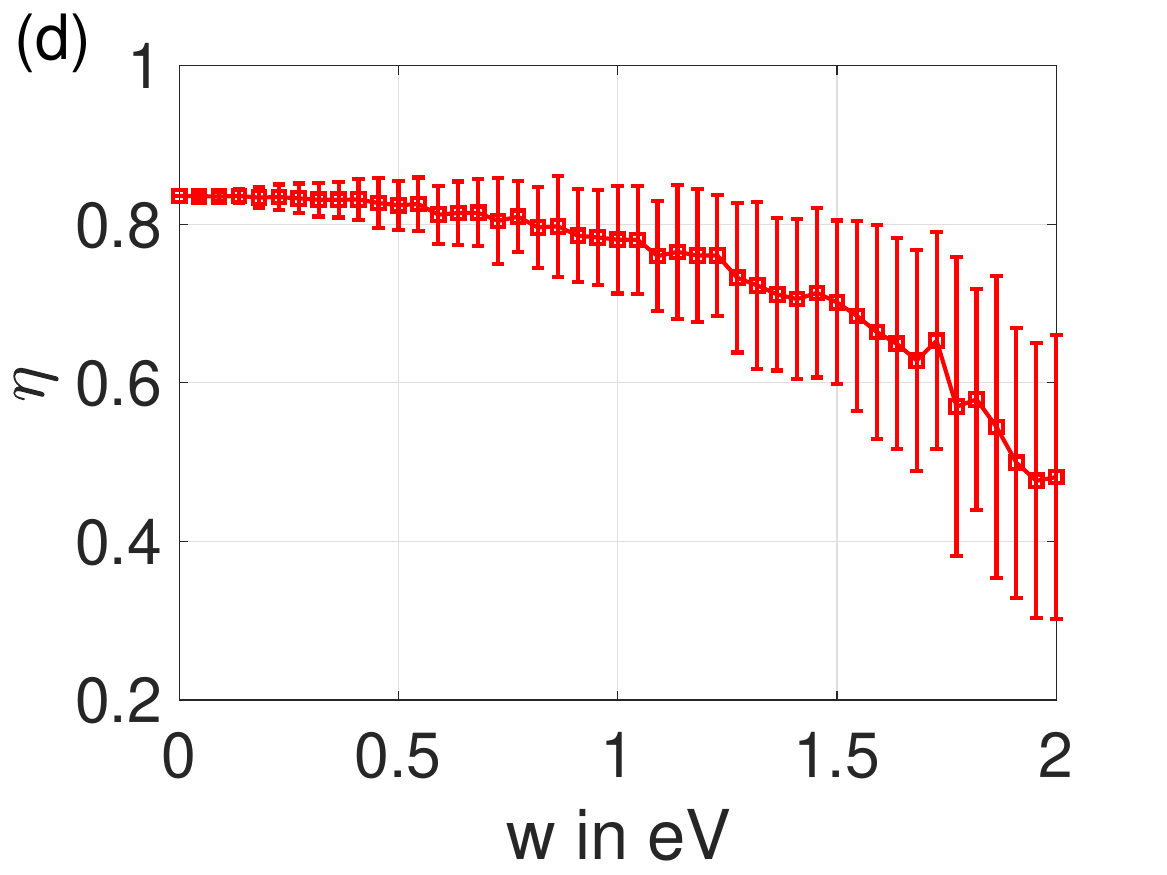}
\caption{(a,c) Conductance through graphene versus disorder strength. (b,d) Valley polarization efficiency versus disorder strength. Both physical quantities are averaged over 100 disorder configurations and the error-bar shows the standard deviation about the mean. Parameters: $L_{gx}=24$, $L_{gy}=24$  $t=1.2~eV$,  $\mu=-1.4~eV$. For (a,b), $\mu_g=0.5eV$, for (c,d) $\mu_g=0.2eV$.  }\label{fig:w}
\end{figure}
 
 {\it Disorder in graphene .-} We introduce on-site disorder in the graphene region, where the disorder potential is randomly distributed within the range \([-w/2, w/2]\). The resulting conductance and valley polarization efficiency are plotted in Fig.~\ref{fig:w}. We observe that the conductance increases with disorder strength for \( \mu_g = 0.5~\text{eV} \) and \( \mu_g = 0.2~\text{eV} \) [see Fig.~\ref{fig:w}(a,c)]. This behavior can be attributed to the low density of states near the charge neutrality point in graphene--introducing disorder enhances the density of states, thereby increasing conductance. However, valley polarization is suppressed by disorder [see Fig.~\ref{fig:w}(b,d)], as impurity scattering leads to mixing between states from the two valleys, reducing the extent of valley polarization.  

{\it Imperfect armchair edges and interface roughness .-}
In realistic experimental setups, the graphene--metal interface is unlikely to be perfectly smooth. We model this interface roughness by introducing non-uniform hopping strengths between graphene and the normal metal. Additionally, imperfections in the armchair edges of graphene are expected and can be captured by introducing on-site disorder near these edges over a finite width along the transverse direction.

\begin{figure}
\includegraphics[width=4cm]{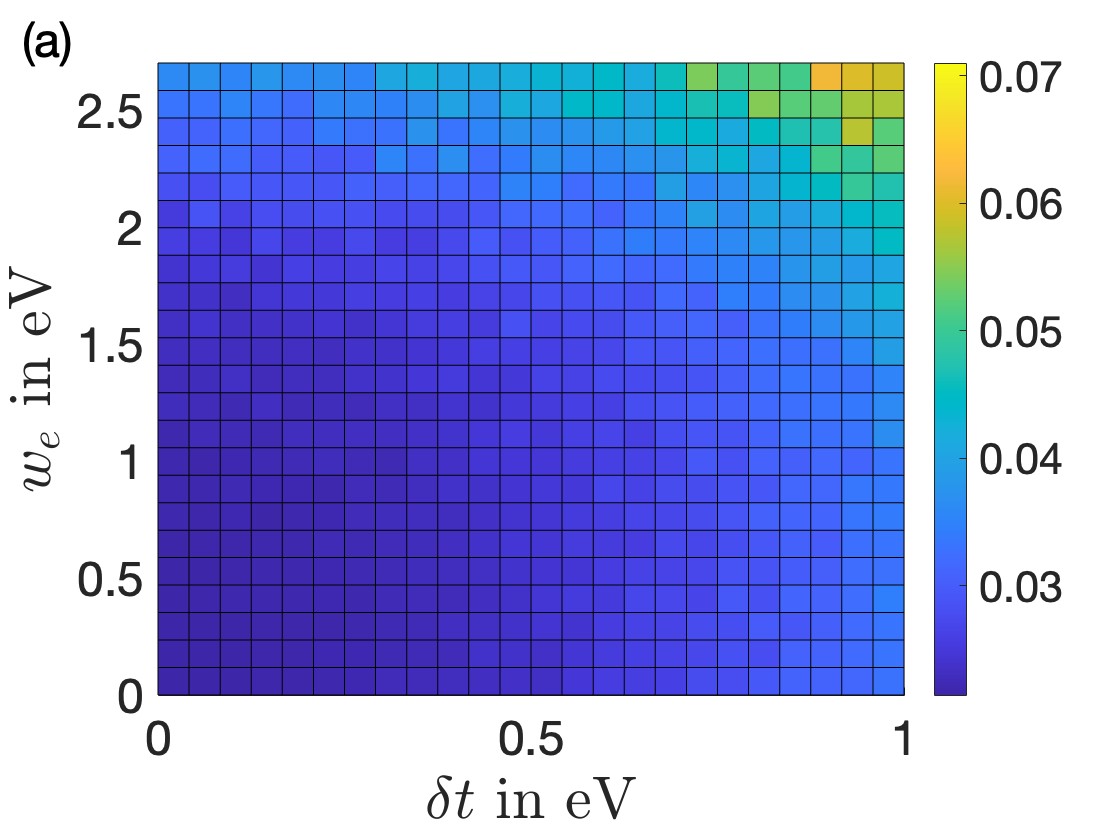}
\includegraphics[width=4cm]{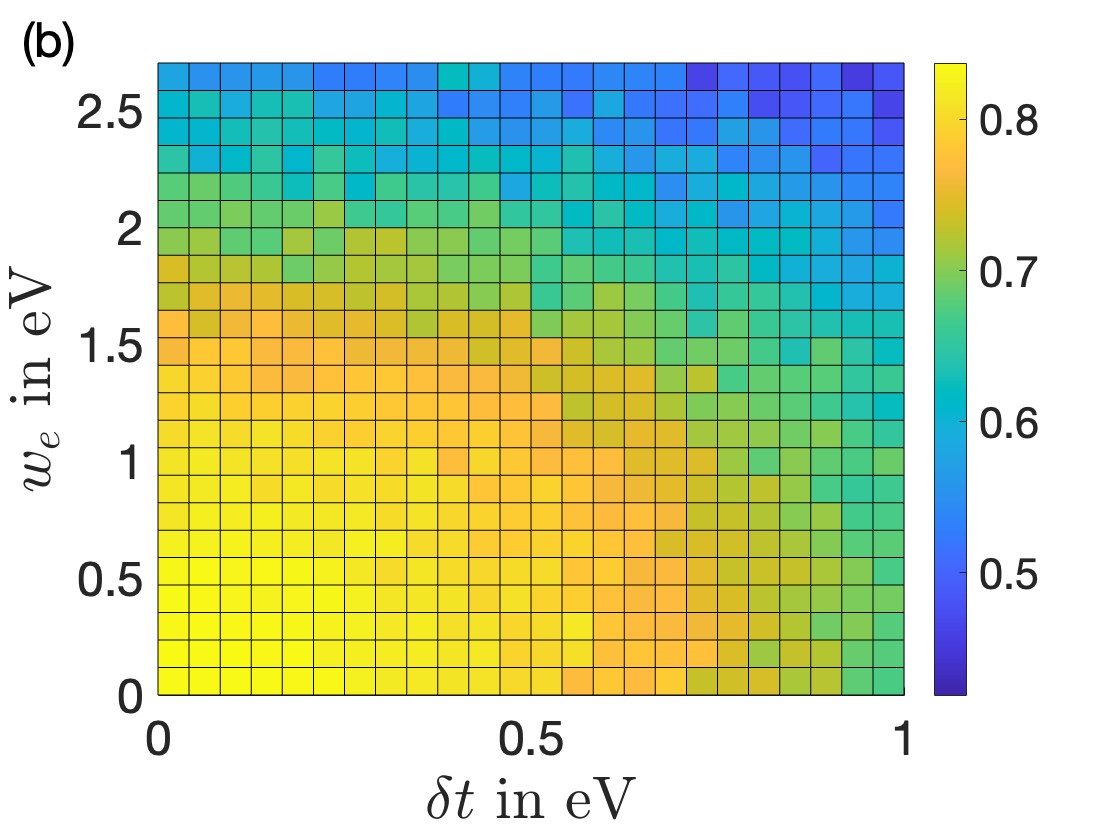}
\includegraphics[width=4cm]{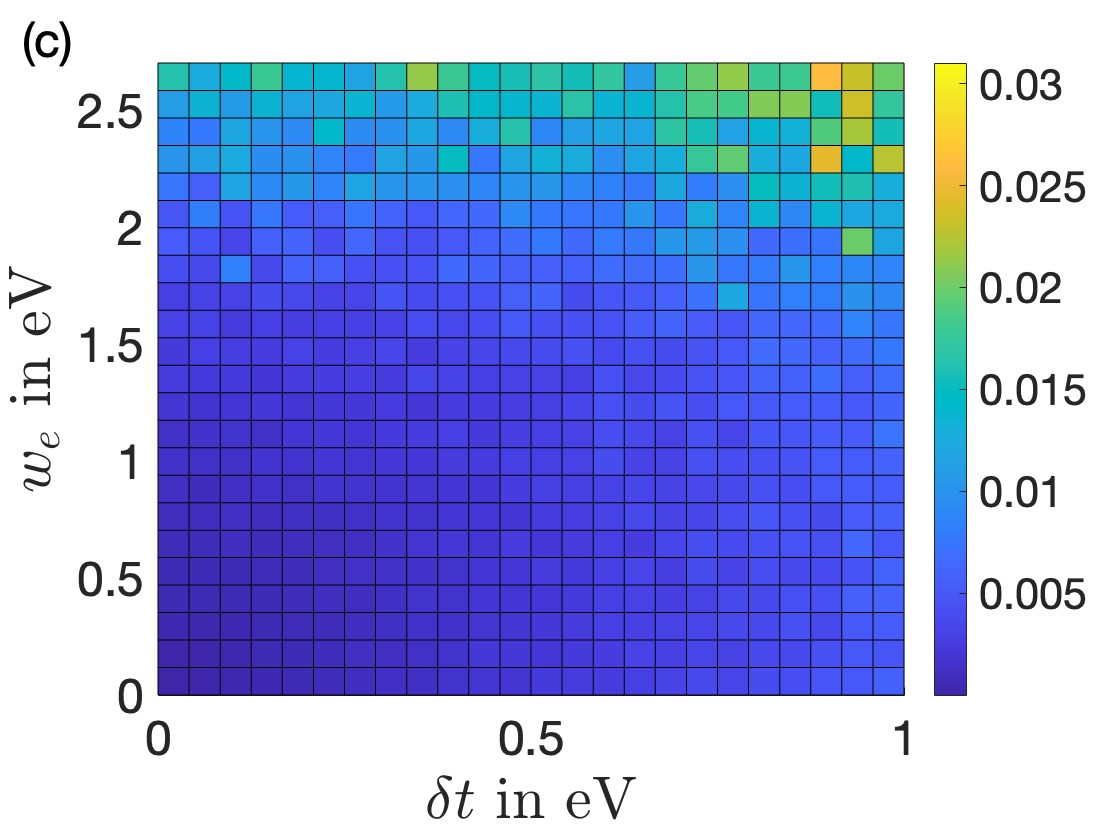}
\includegraphics[width=4cm]{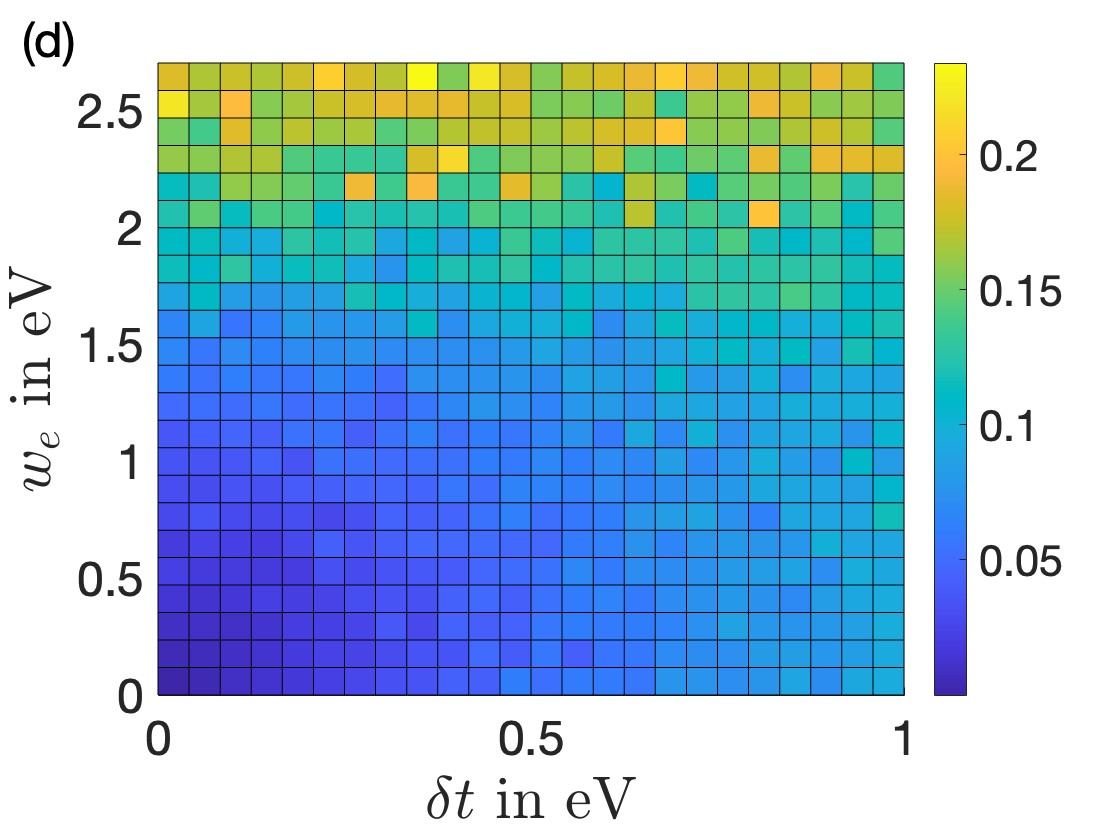}
\caption{Effect of edge disorder and interface roughness on valley polarization and conductance. 
The graphene region has size \( L_{gx} = L_{gy} = 24 \), with on-site disorder of strength \( w_e \) introduced within four lattice sites from the top and bottom armchair edges to model edge imperfections. Interface roughness is implemented by choosing the hopping amplitude \( t'_{l/r,n_y} \) between graphene and normal metal from a uniform random distribution in the range \( [t - \delta t/2, t + \delta t/2] \). 
Panels show the disorder-averaged (a) conductance through graphene and (b) valley polarization efficiency, along with their respective standard deviations (c,d), as functions of \( \delta t \) and \( w_e \). Averages are taken over \( N_c = 50 \) disorder realizations. $\mu_g=0.2 eV$, other parameters are same as in Fig.~\ref{fig:w}.
}\label{fig:rough}
\end{figure}

Specifically, we model the hopping strengths at the left and right interfaces as \( t'_{l/r,n_y} = t + \delta t_{l/r,n_y} \), where \( \delta t_{l/r,n_y} \) is randomly drawn from a uniform distribution in the range \( [-\delta t/2, \delta t/2] \) [see the terms \( H_{LG} \) and \( H_{GR} \) in Supplementary Material]. For a graphene lattice of size \( L_{gx} = L_{gy} = 24 \), we introduce on-site disorder of strength \( w_e \) near the top and bottom armchair edges, i.e., for all sites located within four lattice sites from the transverse boundaries. The disorder potential on these sites is randomly chosen from the range \( [-w_e/2, w_e/2] \).

To quantify the impact of these imperfections, we perform disorder averaging over \( N_c = 50 \) configurations for each pair \( (\delta t, w_e) \). The resulting average conductance through graphene and valley polarization efficiency, along with their standard deviations, are presented in Fig.~\ref{fig:rough}.

It is evident from Fig.~\ref{fig:rough} that the conductance through graphene increases with either type of disorder. The increase in conductance with edge disorder strength $w_e$ can be attributed to an enhanced density of states near the armchair edges. On the other hand, increasing the interface disorder $\delta t$ leads to higher conductance because more of the electrons incident from terminal 1 are scattered by the disordered hopping bonds and are redirected into the graphene sheet, rather than flowing directly into terminal-2 or back into terminal-1. However, both types of disorder: $w_e$ and $\delta t$ lead to a reduction in valley polarization efficiency due to enhanced intervalley mixing caused by scattering. Nonetheless, as seen in Fig.\ref{fig:rough}, the valley polarization remains substantial over a wide range of disorder strengths. For instance, at  $\de t=0.5 eV$ and $w_e = 1.526 eV$, the valley polarization efficiency is $0.6981 \pm  0.0981$. Thus, while a perfectly smooth interface and ideal armchair edges yield optimal valley polarization, small to moderate deviations from these conditions still maintain a high degree of valley polarization.

{\it Discussion and Conclusion.--}
We proposed an all-electrical scheme for achieving valley polarization in graphene, where a graphene sheet is contacted by normal metal electrodes on both sides. When a bias is applied at one terminal and the others are grounded, valley polarization arises if the Fermi wavevector in the metal exceeds half the separation between the \( K \) and \( K' \) valleys in graphene.

We examined how system geometry affects performance. Increasing the graphene width enhances both conductance and valley polarization, while increasing the length initially suppresses conductance and introduces Fabry-Pérot oscillations. Longer lengths also reduce polarization due to increased intervalley scattering.

The effect of disorder was also analyzed. While on-site disorder in graphene enhances conductance near the Dirac point via increased density of states, it concurrently reduces valley polarization due to intervalley mixing. We modeled interface roughness and imperfect armchair edges through randomness in hopping amplitudes and edge-site potentials. Our simulations show that valley polarization remains substantial for moderate deviations from ideal geometry, indicating robustness. Despite challenges in realizing clean interfaces and edges, recent advances in fabrication have demonstrated atomically precise graphene edges~\cite{jia2009,Cai2010}, supporting the experimental feasibility of our model.

In  metals like gold and silver, the Fermi wavevector is around \(1.2~\text{\AA}^{-1}\)~\cite{Ashcroft}, exceeding half the valley separation in graphene (\(\sim 0.85~\text{\AA}^{-1}\)), thus fulfilling the transverse momentum matching criterion. Though these are 3D metals, transport at 2D interfaces is often governed by a few atomic layers. Our square-lattice model effectively captures this using a minimal 2D representation. This is analogous to the use of 1D quantum wires to model reservoirs in mesoscopic transport~\cite{diventra}. By tuning parameters to match realistic Fermi wavevectors, we retain the essential physics required for valley polarization.

In summary, our work offers a robust, tunable, and all-electrical approach to valley polarization in graphene, with promising implications for valleytronic applications.

{\it \noi Acknowledgements .-}
We thank Adhip Agarawala and Manu Jaiswal for illuminating discussions. We thank Manu Jaiswal for comments on the manuscript.  SD thanks Bijay Kumar Sahoo for help with numerics in the early stages of the project. SD and AS thank  Science and Engineering Research Board (now Anusandhan National Research Foundation) - Core Research grant (CRG/2022/004311) for financial support. AS thanks the funding from  Institute of Eminence Professional Development Fund, University of Hyderabad. 

\bibliography{ref_graphene}

\newpage
\setcounter{equation}{0}
\setcounter{figure}{0}
\makeatletter

\onecolumngrid

\begin{center}
{\bf \large Supplementary Material for ``An all-electrical scheme for valley polarization in graphene''} 
\end{center}

\section{Details of calculation}

Figure~1 in the paper illustrates the schematic of the model, where a finite graphene sheet is connected to two normal metals (NMs). These normal metals, modeled as square lattices, have a finite number of sites (\( L_{nx} \)) along the \( x \)-direction, while they extend infinitely along the \( y \)-direction. As shown in the figure, the system consists of four terminals. When a bias is applied to terminal 1 while the other three terminals are grounded, a potential difference develops between terminal 1 and the other terminals. This potential difference drives a current from terminal 1 toward terminal 2, while also inducing currents in graphene toward terminals 3 and 4.  

\begin{figure} [htb]
    \includegraphics[width=10cm]{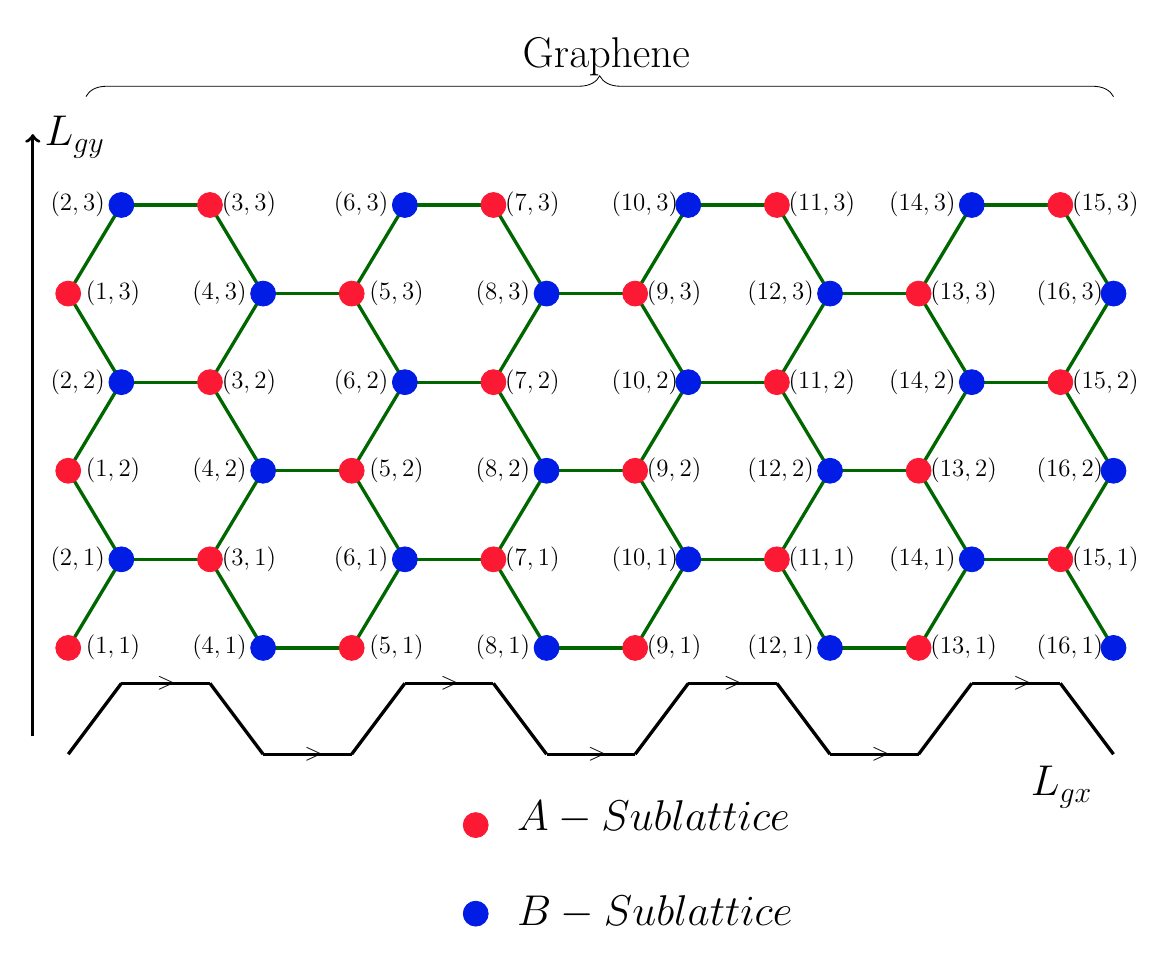}
    \caption{Scheme for labelling the lattice sites in graphene with $L_{gx}=16$, $L_{gy}=3$. }
    \label{fig:graphene}
\end{figure}

The Hamiltonian for the setup is given by:  

\bea
  H &=& H_L + H_{LG} + H_G + H_{GR} + H_R, {\rm~~~ where,} \nn \\ 
   H_{LG}&=& -\sum_{n_y=1}^{L_{gy}} t'_{l,n_y} [c^{\dagger}_{L_{nx},n_y}d_{1,n_y}+{\rm h.c.}],\nn \\
 H_{GR}&=&- \sum_{n_y=1}^{L_{gy}} t'_{r,n_y}[d^{\dagger}_{L_{gx},n_y}c_{L_{nx}+1,n_y}+{\rm h.c.}]\nn \eea
  
\bea 
H_L &=&-t  \sum_{n_y=-\infty}^{\infty}\Big[ \sum_{n_x=1}^{L_{nx}-1} [(c^{\dagger}_{n_x+1,n_y}c_{n_x,n_y}+{ \rm h.c.}) ] 
+ \sum_{n_x=1}^{L_{nx}}[(c^{\dagger}_{n_x,n_y-1}c_{n_x,n_y}+{\rm h.c.})]\Big]    - \mu_{n}\sum_{n_x=1}^{L_{nx}}\sum_{n_y=-\infty}^{\infty} c^{\dagger}_{n_x,n_y}c_{n_x,n_y} \nn   
\eea 
\bea
H_R &=&-t  \sum_{n_y=-\infty}^{\infty}\Big[ \sum_{n_x=L_{nx}+1}^{2L_{nx}-1} [(c^{\dagger}_{n_x+1,n_y}c_{n_x,n_y}+{ \rm h.c.}) ]  
+ \sum_{n_x=L_{nx}+1}^{2L_{nx}}[(c^{\dagger}_{n_x,n_y-1}c_{n_x,n_y}+{\rm h.c.})]\Big]     - \mu_{n}\sum_{n_x=L_{nx}+1}^{2L_{nx}}\sum_{n_y=-\infty}^{\infty} c^{\dagger}_{n_x,n_y}c_{n_x,n_y} \nn
 \eea
    
    \bea 
    H_G &=& -\ga \sum_{n_x=1}^{L_{gx}-1} \sum_{n_y=1}^{L_{gy}}[d^{\dagger}_{n_x+1,n_y}d_{n_x,n_y} + {\rm h.c.}] 
    -\ga\sum'_{n_x=4n+2}\sum_{n_y=1}^{L_{gy}-1}[d^{\dagger}_{n_x-1,n_y+1}d_{n_x,n_y} + {\rm h.c.}] \nn \\ 
  && -\ga\sum'_{n_x=4n+3}\sum_{n_y=1}^{L_{gy}-1}[d^{\dagger}_{n_x+1,n_y+1}d_{n_x,n_y} + {\rm h.c.}] 
   -\mu_g\sum_{n_x=1}^{L_{gx}}\sum_{n_y=1} ^{L_{gy}}d^{\dagger}_{n_x,n_y}d_{n_x,n_y}, \label{eq:ham}
    \eea
where  $c_{n_x,n_y}$ annihilates an electron  at site $(n_x, n_y)$ in normal metal, $d_{n_x,n_y}$ annihilates an electron on site $(n_x,n_y)$ in graphene, $t$ is the hopping strength in the normal metal, $t'_{l/r,n_y}$ is the hopping strength at the junction of the normal metal and graphene at $n_y$ on left/right, and $\mu_{n}$ is the chemical potential in the normal metal, $\ga$ is the hopping strength in graphene, $\mu_g$ is the chemical potential in graphene, $\sum'_{n_x=4n+m}$ means the summation is over all integers between $1$ to $L_{gx}$ having the form $(4n+m)$, where $n$ is a nonnegative integer and $m=2, 3$. We take graphene to be composed of spinless electrons. We set $t'_{l/r,n_y}=t$ in numerical calculations for producing results leading to Figures~3,~4,~5,~6 of the main paper.  The labelling of sites on the graphene lattice follows the scheme shown in Fig.~\ref{fig:graphene} for lattice of size $(L_{gx},L_{gy})=(16,3)$. 

In each of the four terminals of the normal metal, the modes are described by the wavenumber \( k_m \) in the \( y \)-direction, which satisfies the dispersion relation  
\beq  
E = -2t \cos{(k_m b)} + \epsilon_m - \mu_n,  
\eeq  
for \( m = 1,2,\dots,L_{nx} \), where \( b \) is the lattice constant in the normal metal. Here, \( \epsilon_m \) are the eigenenergies of a one-dimensional Hamiltonian along the \( x \)-direction with open boundary conditions, having \( L_{nx} \) sites and a hopping strength of \( t \). As a result, each of the four terminals supports \( L_{nx} \) transport channels.  

The column vector \( v_m \), of size \( L_{nx} \times 1 \), represents the eigenstate of the one-dimensional Hamiltonian corresponding to the eigenenergy \( \epsilon_m \). At a given energy \( E \), the wavenumber \( k_m \) may be real or complex. An electron can be incident from channel \( m_0 \) of terminal 1 only if \( k_{m_0} \) is real.  

The scattering eigenfunction corresponding to an electron with energy \( E \) incident on the system is expressed as  
\beq  
|\psi\rangle = \sum_{n_x,n_y} \psi_{n_x,n_y,n} |n_x,n_y,n\rangle + \sum_{n_x,n_y} \psi_{n_x,n_y,g} |n_x,n_y,g\rangle.  
\eeq  
Here, \( \psi_{n_x,n_y,n} \) can be grouped into column vectors \( \psi_{n_y,n_t} \) with \( L_{nx} \) entries, where the \( n_x \)-th entry is equal to \( \psi_{n_x,n_y,n} \) in terminals \( n_t = 1,2 \), and the \( n_x \)-th entry is equal to \( \psi_{L_{nx}+n_x,n_y,n} \) in terminals \( n_t = 3,4 \). The wave function \( \psi_{n_y,n_t} \) takes the form:  

\bea 
\psi_{n_y,1} &=&  v_{m_0}~e^{ik_{m_0}n_yb} + \sum_{m=1}^{L_{nx}} r_{m,m_0} v_m e^{-i k_{m} n_y b},\nn \\&& ~~~~~~~~~~~~~~{\rm for~~}n_y \le 0, \nn \\ 
\psi_{n_y,2} &=& \sum_{m=1}^{L_{nx}} t_{m,m_0,2} v_me^{ik_mn_yb}, ~~{\rm for ~~} n_y >L_{gy}, \nn \\ 
\psi_{n_y,3} &=& \sum_{m=1}^{L_{nx}} t_{m,m_0,3} v_me^{ik_mn_yb}, ~~{\rm for ~~} n_y >L_{gy}, \nn \\ 
\psi_{n_y,4} &=& \sum_{m=1}^{L_{nx}} t_{m,m_0,4} v_me^{-ik_mn_yb}, ~~{\rm for ~~} n_y \le 0. \nn \\ 
\eea

At a given energy $E$, $k_m$ is chosen to be positive if real and has positive imaginary part if complex.  The scattering coefficients $r_{m,m_0}$, $t_{m,m_0,j}$ (for $j=2,3,4$) and the wavefunctions $\psi_{n_x,n_y,g}$ can be determined using the Schr\"odinger wave equation $H|\psi\ra=E|\psi\ra$. 

     The differential conductivity $G_j=dI_j/dV$, the differential ratio of the current in terminal $j$ to the voltage bias $V$ applied in terminal-1 is given by: 
     \bea
G_{j} &= &\frac{e^2}{h}\mathlarger{\mathlarger{\sum'}}_{m_0}\frac{1}{\sin{k_{m_0}a}}\mathlarger{\mathlarger{\sum'}}_{m}{|t_{m,m_0}^{j}|^2\sin{k_{m}a}},\\&&~~~~~~~~~~~~~~~~~(\rm{for}~  j=2,3,4), \nn\\ G_{1} &=& \frac{e^2}{h}\mathlarger{\mathlarger{\sum'}}_{m_0}\Big[1-\mathlarger{\mathlarger{\sum'}}_{m}\frac{|r_{m,m_0}|^2\sin{k_{m}a}}{\sin{k_{m_0}a}}\Big], \nn \\ &&
\eea
where the primed summation over $m$ means that the summation is done over all the values of $m$ for which $k_m$ is real. The conductances obey the conservation condition $G_1=G_2+G_3-G_4$. The conductance through graphene is given by $G_g=G_1-G_2=G_3-G_4$.
  
  To characterize the valley polarization, we consider the momentum eigenstates of the graphene lattice, denoted as $\phi_{\vec{k}}$, which have the same dimensions as the graphene region in the setup and satisfy periodic boundary conditions within the first Brillouin zone. The states near the $K$-point are classified as belonging to the $K$ valley, while those near the $K'$-point are assigned to the $K'$ valley.  

Let $\psi(m_0,E)$ represent a current-carrying eigenstate of the scattering problem corresponding to an electron incident in the $m_0$-th transport channel at energy $E$. The graphene component of this scattering eigenfunction, normalized over the graphene region, is denoted as $\psi^g(m_0,E)$. If $G_g(m_0,E)$ is the contribution of the $m_0$-th channel to the total conductance through graphene, then the contribution of a particular momentum state $\vec{k}$ in graphene to this conductance is given by  
\[
G_g(m_0,E) |\phi_{\vec{k}}^{g \dagger} \cdot \psi^g(m_0,E)|^2.
\]

 The valley polarised conductance which is defined as the difference between conductance contributions from states belonging to $K$ and $K'$ valleys is then given by 
 \bea 
 G^{g,V} &=& \sum' _{m_0}G_g(m_0,E) \cdot\big[\sum_{\vec k\in K}|\phi_{\vec k}^{\dagger} \cdot \psi^g(m_0,E)|^2 - \sum_{\vec k\in K'}|\phi_{\vec k}^{\dagger} \cdot \psi^g(m_0,E)|^2\big] 
 \eea
 
 The valley polarization efficiency $\eta$ defined as the ratio of difference between the contributions of $K$ and $K'$ valleys to conductance through graphene to the mean of the contributions of $K$ and $K'$ valleys to conductance through graphene is given by 
\bea  \eta &=& \f{2 \sum' _{m_0}G_g(m_0,E) \cdot\big[\sum_{\vec k\in K}|\phi_{\vec k}^{\dagger} \cdot \psi^g(m_0,E)|^2 - \sum_{\vec k\in K'}|\phi_{\vec k}^{\dagger} \cdot \psi^g(m_0,E)|^2\big] }{ \sum' _{m_0}G_g(m_0,E) \cdot\big[\sum_{\vec k\in K}|\phi_{\vec k}^{\dagger} \cdot \psi^g(m_0,E)|^2 + \sum_{\vec k\in K'}|\phi_{\vec k}^{\dagger} \cdot \psi^g(m_0,E)|^2\big] } \eea

\section{Dependence of $G_g$ on $L_{gx}$}\label{sec:app2}
\begin{figure}
 \includegraphics[width=6cm]{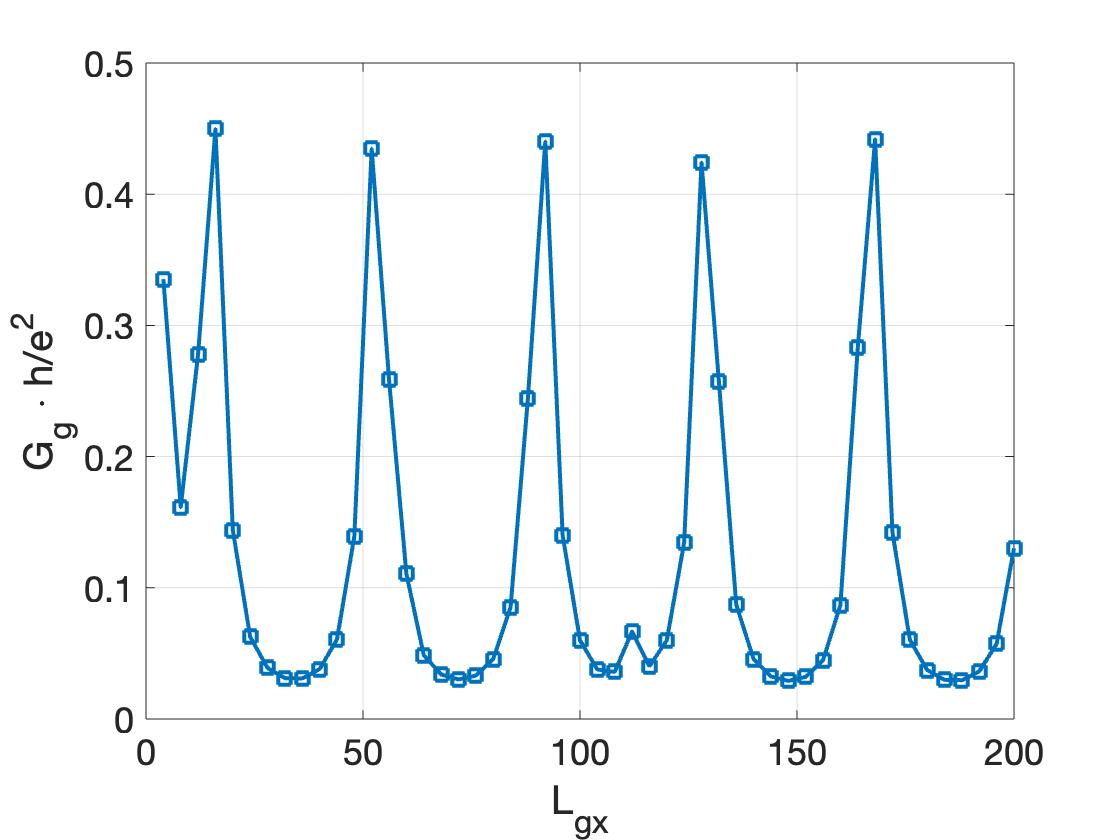}
 \caption{Conductance through graphene as a function of \( L_{gx} \) for a fixed transverse width \( L_{gy} = 12 \). Compared to Fig.~4 of the main paper, where \( L_{gy} = 24 \), the oscillations are more regular here due to fewer transverse modes contributing to transport. With fewer channels, the interference pattern is simpler and more periodic, illustrating clearer Fabry--Pérot resonances.}\label{fig:small_Lgy}
 \end{figure}

We show in Fig.~\ref{fig:small_Lgy} that the conductance through graphene as a function of \( L_{gx} \) exhibits more regular oscillations when the transverse width is reduced to \( L_{gy} = 12 \), keeping all other parameters the same as in Fig.~4 of the main paper. This increased regularity arises because a smaller \( L_{gy} \) limits the number of transverse modes (or channels) participating in transport. When fewer channels contribute, the interference pattern is dominated by a smaller set of longitudinal wave vectors, leading to more regular Fabry--Pérot-like oscillations. In contrast, as more channels participate (i.e., for larger \( L_{gy} \)), the superposition of multiple interference conditions results in irregular oscillations due to differing resonance conditions across channels. However, reducing  $L_{gy}$  also diminishes both the conductance and the valley polarization, as fewer channels contribute to transport and the transverse momentum matching becomes less efficient. 

\end{document}